\documentclass[journal=jpcafh,layout=twocolumn,manuscript=article]{achemso}
\setkeys{acs}{doi = true, chaptertitle=true}
\usepackage{hyperref}
\providecommand{\doi}[1]{}
\renewcommand{\doi}[1]{%
  \href{https://doi.org/#1}{\nolinkurl{#1}}%
}

\usepackage[version=3]{mhchem} 
\usepackage{booktabs}
\usepackage{graphicx}
\usepackage[table,xcdraw]{xcolor}
\usepackage{chemformula} 
\usepackage[T1]{fontenc} 
\usepackage{lscape}

\author{Margarita Shepelenko}
\affiliation[Weizmann Institute of Science ]{Department of Molecular Chemistry and Materials Science, Weizmann Institute of Science, 7610001 Re\d{h}ovot, Israel}
\altaffiliation{Equally contributing first authors}
\author{Golokesh Santra}
\affiliation[Weizmann Institute of Science ]{Department of Molecular Chemistry and Materials Science, Weizmann Institute of Science, 7610001 Re\d{h}ovot, Israel}
\alsoaffiliation[PsiQuantum]{Present address: PsiQuantum, Ltd., Bristol, UK}
\altaffiliation{Equally contributing first authors}
\author{Emmanouil Semidalas}
\affiliation[Weizmann Institute of Science ]{Department of Molecular Chemistry and Materials Science, Weizmann Institute of Science, 7610001 Re\d{h}ovot, Israel}
\alsoaffiliation[Athens]{Present address: Department of Chemistry, National and Kapodistrian University of Athens, Zografou 15772, Athens, Greece}
\author{Jan~M.~L. Martin}
\affiliation[Weizmann Institute of Science ]{Department of Molecular Chemistry and Materials Science, Weizmann Institute of Science, 7610001 Re\d{h}ovot, Israel}
\email{gershom@weizmann.ac.il}
\phone{+972-8-9342533}
\fax{+972-8-9343029}

\usepackage{xcolor}
\usepackage{etoolbox}
\usepackage{verbatim}

\title[post-CCSD(T) in water clusters]
  {Water clusters beyond valence CCSD(T): Isomers of \ce{(H2O)6} and \ce{(H2O)20} with special reference to BSSE}
\begin{document}


\begin{tocentry}
\includegraphics[width=2in]{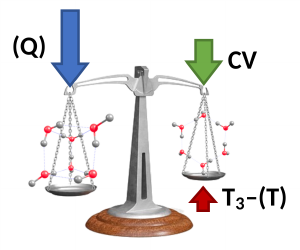}
\end{tocentry}

\begin{abstract}
Benchmark calculations on noncovalent interactions typically exclude correlation effects beyond valence CCSD(T) owing to their steep computational cost scaling. In this work, we consider their importance for water clusters, specifically, eight isomers of \ce{(H2O)6} and four Wales-Hodges isomers of \ce{(H2O)20}. 
Basis set superposition error (BSSE) was treated by the Valiron-Mayer approach. Higher-order connected triples, $T_3$--(T), reduce dissociation energies of the latter by about 0.4 kcal/mol, but connected quadruples \emph{increase} them by about twice that, 0.8 kcal/mol. In general, connected quadruples favor more compact isomers over more `spread-out' ones, while higher-order triples effects are relatively constant across isomers. We also consider additional small effects for balance: scalar relativistics reduce binding in \ce{(H2O)20} by ca. 0.4 kcal/mol, which fortuitously is (somewhat over)compensated by the ca. 0.6 kcal/mol diagonal Born-Oppenheimer correction. Core-valence correlation has the greatest impact, at 1.3--1.4 kcal/mol for the various icosamer structures. While core-valence correlation tends to favor more open structures, (Q) is roughly proportional to the number of hydrogen bonds and thus favors more compact structures.  Overall, when BSSE is properly accounted for, two-body contributions account for over 95\% of contributions beyond valence CCSD(T). 
\end{abstract}

\section{Introduction}

Noncovalent interactions are ubiquitous in supramolecular chemistry; chemical biology, e.g., molecular recognition and secondary structure of biomolecules; physical chemistry of the liquid state; and more. Among noncovalently bound aggregates, water clusters hold a special position, in attempts to understand the unique physicochemical properties of `life's solvent' (see the introductions to Refs.\citenum{Paesani2020,XantheasAtlas,Saykally2001,jmlm272,Bowman2021,Tschumper2014,Burke2023}).

The `gold standard' for benchmarking noncovalent interactions has for some time been valence CCSD(T) (coupled cluster with all singles and doubles and quasiperturbative connected triples\cite{Rag89,Wat93}) extrapolated to the complete basis set limit by various stratagems (see, e.g., the recent review by Karton\cite{Karton2022}). For larger molecules, localized coupled cluster approaches are increasingly employed. 

Owing to the very steep computational cost scaling of higher-order electron correlation methods — for instance, CCSDT\cite{Noga1986} and CCSDT(Q)\cite{mrcc8} will scale as the eighth and ninth (!) power, respectively, of the system size — there have been few attempts to determine post-CCSD(T) corrections, and these have mostly focused on small species. Boese\cite{Boese2013} considered hydrogen bonding of diatomics, triatomics, and tetratomics; Hobza and coworkers\cite{A24,A24bis} focused on their A24 benchmark; while recently Karton and Martin\cite{jmlm298} investigated the benzene dimer. Patkowski and coworkers\cite{Smith2014PostCCSDT} investigated post-CCSD(T) corrections for small noncovalent dimers and concluded that basis sets as large as aug-cc-pVTZ\cite{Dun89,Kendall1992} may be necessary for adequate convergence.

It is well established in computational thermochemistry (e.g., Refs.\citenum{jmlm200,heat3,jmlm273,Karton2022}) that for molecules dominated by dynamical (rather than static\cite{Hollett2011}) correlation, CCSD(T) is a Pauling point owing to error cancellation between neglect of, usually antibonding, higher-order triple excitations CCSDT -- CCSD(T), and universally bonding connected quadruple excitations.

Application of CCSDT(Q) to problems like the L7 (seven large noncovalent complexes) benchmark\cite{L7} is not realistic at present ---  and yet here, discrepancies between approximate CCSD(T) and quantum Monte Carlo approaches\cite{Hamdani2021} prompt the question whether the error cancellation in CCSD(T) is starting to break down. This has led to several studies (e.g., Refs.\citenum{Grueneis2025,jmlm335}) showing that in $\pi$-stacks nontrivial post-CCSD(T) corrections exist -- while the higher-order triples $E$[CCSDT]$-E$[CCSD(T)] narrow the gap with quantum Monte Carlo,\cite{Grueneis2025} connected quadruple excitations  $E$[CCSDT(Q)]$-E$[CCSDT] reopen the gap, leaving CCSD(T) in better agreement with CCSDT(Q) than with CCSDT. Recent studies\cite{lambie2026nodal,fanta2025why} show how flaws in fixed-node QMC are likely at the root of the problem.

We have found\cite{jmlm336} much smaller post-CCSD(T) effects in hydrogen-bonded dimers from the S66 dataset\cite{HobzaS66}. However, is this still the case in neutral water clusters (H$_2$O)$_n$, given how many such interactions they involve? As it happens, a many-body expansion (MBE, Ref.\citenum{PolishMBEpaper} and references therein) of their energies converges fairly rapidly\cite{Paesani2016,Herbert2017,jmlm272}, and we saw indications\cite{jmlm272} that the further up the electron correlation `ladder' one climbs, the faster the MBEs for the successive contributions converge. This offers the tantalizing prospect that post-CCSD(T) contributions of fairly large water clusters may be amenable to evaluation through MBE, and that thus insights on the nature of post-CCSD(T) contributions to noncovalent interactions may be obtained. 

\begin{figure*}[t]
\centering
\includegraphics[width=\textwidth]{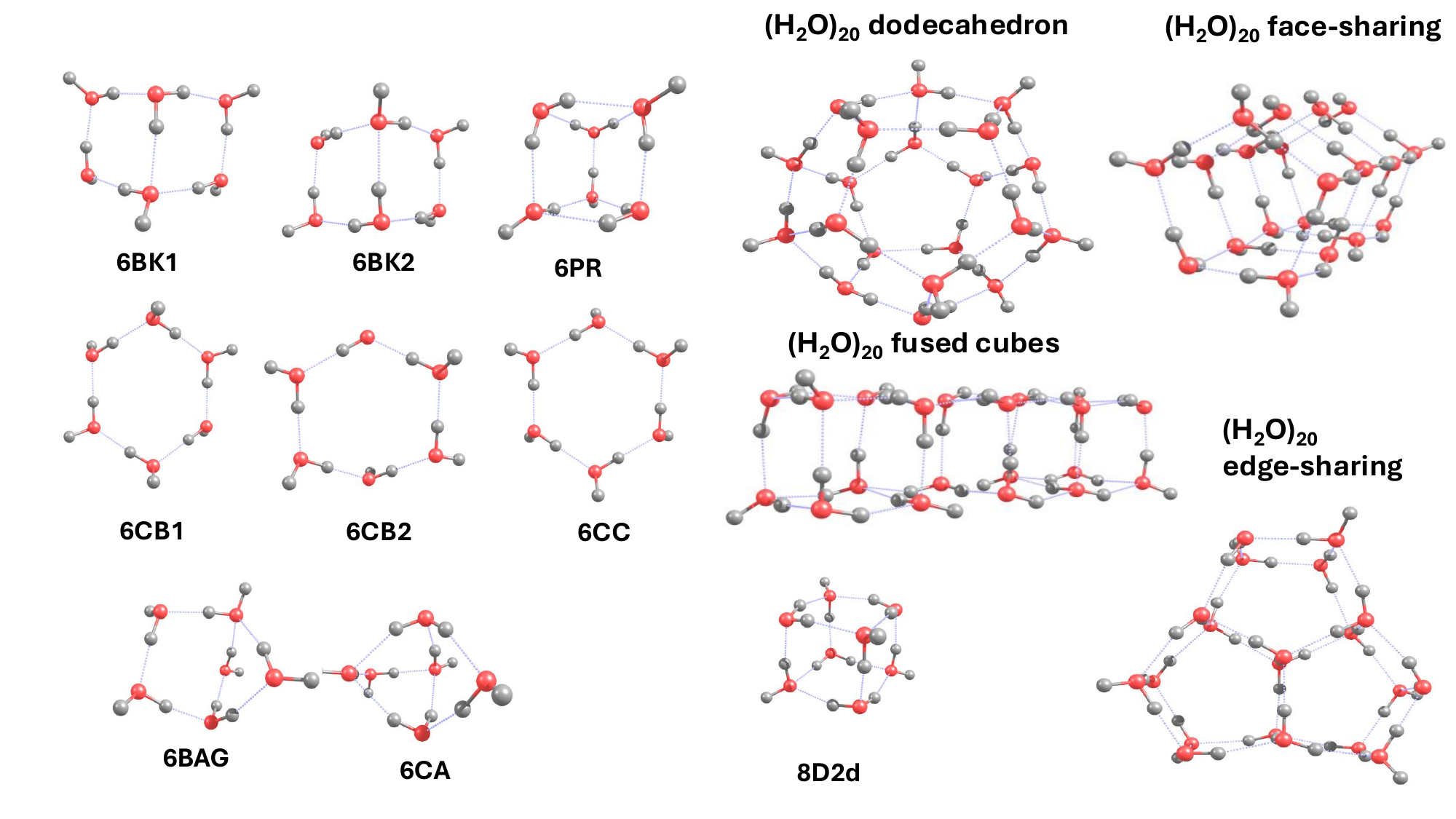}
\caption{Water cluster isomers considered in this study}
\label{fig:waterclusters}
\end{figure*}

We have explored this hypothesis for two systems (Figure~\ref{fig:waterclusters}). The first are the eight isomers of \ce{(H2O)6}, which has been extensively studied (e.g, Refs.\citenum{Temelso2011,Xantheas2013,jmlm272} and references therein) as the smallest cluster that has a three-dimensional (rather than cyclic) global minimum structure. The second are the four \ce{(H2O)20} isomers that occur in the WATER27 dataset\cite{WATER27}  --- the said isomers represent the four Wales-Hodges\cite{Wales1998,Xantheas2004} structure classes: dodecahedron, fused-cube (a.k.a., box-kite, stacked cubes), face-sharing (stacked pentagons), and edge-shared.

For the sake of perspective, we have also evaluated other contributions that might conceivably affect cluster binding energies in the same energy range, such as core-valence correlation, scalar relativistics, and the diagonal Born-Oppenheimer correction.

\section{Methods}

All CCSD(T) or lower-level calculations reported here were carried out using MOLPRO 2022.3\cite{Molpro2020}, aside from the diagonal Born-Oppenheimer corrections (DBOC), which were obtained using the implementation\cite{Gauss2006} in CFOUR 2.1\cite{Cfour2020}. CCSDT, CCSDT(Q), and CCSDTQ results were likewise obtained using CFOUR (specifically the NCC module\cite{Matthews2015} developed by Devin A. Matthews and coworkers on top of the TBLIS tensor library\cite{TBLIS}), while CCSDTQ(5) calculations were carried out by means of MRCC\cite{MRCC,mrcc63}. The Weizmann Institute's Faculty of Chemistry HPC facility CHEMFARM was used throughout. As in Ref.\citenum{jmlm272}, structures for the eight \ce{(H2O)6} isomers and for the lowest isomer of \ce{(H2O)8} were downloaded from the BEGDB noncovalent interactions database\cite{BEGDB,Temelso2011} and used as is, while those for the four \ce{(H2O)20} isomers were taken from the WATER27 benchmark (also part of the GMTKN55\cite{GMTKN55}~---~general main-group thermochemistry, kinetics, and noncovalent interactions --- benchmark suite).

Basis sets used for the post-CCSD(T) and DBOC correction are the Dunning correlation consistent cc-pV$n$Z ($n$=D, T, Q) basis sets\cite{Dun89}, as well as their (diffuse function-)augmented variants\cite{Kendall1992} aug-cc-pV$n$Z. The addition of diffuse functions not to hydrogen but only to heavier and more electronegative elements is variously denoted in the literature as aug$'$-cc-pV$n$Z (following Janet E. Del Bene\cite{DelBene1993}), haV$n$Z (following Pavel Hobza\cite{Rezac2016}), and as the `calendar basis set' jul-cc-pV$n$Z (following Papajak and Truhlar\cite{PapajakTruhlar}). In addition, we use suffixes to indicate angular moment truncation: specifically, haVTZ(f,p) refers to cc-pVTZ with $d$ functions omitted on hydrogen and full aug-cc-pVTZ on oxygen, and haVQZ(g,d) to cc-pVQZ minus the $f$ functions on hydrogen and full aug-cc-pVQZ on oxygen.

For the core-valence calculations, we employed aug-cc-pwCV$n$Z (n=T, Q, 5) basis sets\cite{Peterson2002} on oxygen combined with un-augmented cc-pV$n$Z on hydrogen, this combo being abbreviated as hawCV$n$Z. (For a detailed discussion of basis set convergence in this contribution, see Ref.\citenum{jmlm281}. For this term, simple Helgaker $L^{-3}$ extrapolation\cite{Halkier1998} was applied.) 

Scalar relativistic corrections were obtained using the 2nd-order Douglas-Kroll-Hess approximation\cite{Reiher2012} with a fully decontracted heavy-aug-cc-pVTZ basis set (haVTZuc), that is, aug-cc-pVTZuc on oxygen and cc-pVTZuc on hydrogen.

The `brace' notation, e.g., V\{D,T\}Z, refers to extrapolation from cc-pVDZ and cc-pVTZ basis sets. Post-CCSD(T) extrapolation parameters were taken from Table 3 of Karton\cite{Karton2020}. (For a discussion of the equivalence relationships between various two-point extrapolations, see Ref.\citenum{jmlm280}.)

Basis set superposition error (BSSE) was evaluated by the Valiron-Mayer function counterpoise (VMFC) method\cite{valiron1997hierarchy} implemented in a Python script. In a nutshell, VMFC is an adaptation of the standard Boys-Bernardi counterpoise method\cite{Boys1970} to $n$-body expansions: for any individual $n$-body dissociation energy term in an $N$-mer, the energies of the monomers are evaluated in the full $n$-body basis set (i.e., Wells and Wilson's site-site function counterpoise, SSFC, method\cite{Wells1983a}). For $n$=2, VMFC is equivalent to PAFC (pairwise-additive function counterpoise\cite{Wells1983a}); for $n\geq3$, all lower subclusters go into the $n$-mer basis;for $n=N$, the cumulative total dissociation energy becomes equivalent to SSFC. While we recently showed\cite{jmlm333} that post-CCSD(T) differential basis set superposition error is quite weak in noncovalent dimers, this cannot be taken for granted in clusters with many such interactions, so we shall consider both `noCP' and `VMFC' dissociation energies. At the complete basis set limit as estimated by extrapolation, `noCP' and `VMFC' limits should be identical (or merely very close, if the basis sets used in the extrapolation are adequate).

For the avoidance of doubt, all `dissociation energies' in the present paper are frozen-monomer dissociation energies, i.e., the monomer energies are evaluated at their geometries in the cluster. That is to say, positive contributions are attractive, and negative ones repulsive.

\section{Results and discussion}

\subsection{Post-CCSD(T) effects in \ce{(H2O)6} isomers and the global minimum of \ce{(H2O)8} }

\begin{table*}[h!]

\caption{Dissociation energy contributions for (H$_2$O)$_{6}$ isomers (kcal/mol)\label{tab:water6} and (H$_2$O)$_{8}$}
\centering
\resizebox{0.99\textwidth}{!}{
\begin{tabular}{lrrrrrrrrr}\hline
& \multicolumn{8}{c}{\ce{(H2O)6}}&\ce{(H2O)8}\\
& prism & cage & book1 & book2 & bag & cyclic chair & boat1 & boat2 & $D_{2d}$\\
& Iso1 PR & Iso2 CAG & Iso3 BK1 &Iso4 BK2 & Iso5 BAG & Iso6 CC & Iso7 CB1 & Iso8 CB2 & cube\\
\hline
&\multicolumn{8}{c}{valence CCSD(F12*)/CBS +(T) limit for perspective only}\\
Ref.\citenum{jmlm272}& 48.88 & 48.53 & 48.10 & 47.83 & 47.37 & 46.98 & 45.98 & 45.88\\
\hline

$T_3-$(T) VTZ(f,p) 2-body noCP& -0.098 & -0.100 & -0.099 & -0.099 & -0.100 & -0.096	& -0.095 &-0.095\\
$T_3-$(T) VTZ(f,p) 2-body VMFC&-0.084 & -0.087 & -0.087 & -0.088 & -0.087 & -0.086	& -0.085 &-0.084\\
$T_3-$(T) VQZ(g,d) 2-body noCP& -0.121 & & & & & -0.119 \\
$T_3-$(T) VQZ(g,d) 2-body VMFC& -0.088 & & & & & -0.092 \\
$T_3-$(T) haVTZ(f,p) 2-body noCP& -0.072 & -0.074& -0.077& -0.077 & -0.076 & -0.079 & -0.076 & -0.076&-0.109\\
$T_3-$(T) haVTZ(f,p) 2-body VMFC& -0.060 & -0.063& -0.068& -0.068 & -0.066 & -0.071 & -0.068 & -0.068&-0.093\\
$T_3-$(T) haVQZ(g,d) 2-body noCP& -0.088 &-0.092 &-0.095 &-0.095 &-0.094& -0.096 &-0.093&-0.093&-0.137 \\
$T_3-$(T) haVQZ(g,d) 2-body VMFC& -0.067 &-0.070 &-0.075 &-0.075 &-0.074    & -0.078 &-0.076&-0.075&-0.104 \\
$T_3-$(T) haV\{T,Q\}Z 2-body noCP& -0.10&-0.11&-0.11&-0.11&-0.11&-0.11&-0.11&-0.11&-0.16\\
$T_3-$(T) haV\{T,Q\}Z 2-body VMFC&-0.07&-0.08&-0.08&-0.08&-0.08&-0.08&-0.08&-0.08&-0.11\\
$T_3-$(T) best estimate & -0.09$\pm$0.01 & -0.09$\pm$0.02 & -0.10$\pm$0.02 &-0.10$\pm$0.02 &-0.10$\pm$0.01 &-0.10$\pm$0.01 &-0.10$\pm$0.01&-0.10$\pm$0.01 &-0.14$\pm$0.02 \\
$T_3-$(T) VTZ(d,p) 3-body noCP& 0.021 & 0.018 & 0.016 & 0.016 & 0.016  & 0.012 & 0.012  & 0.012 &  0.039\\
$T_3-$(T) VTZ(d,p) 3-body VMFC & -0.002 & & & & & +0.006 &&&0.000\\
$T_3-$(T) haVTZ(f,p) 3-body noCP& +0.002 & & & & & +0.008 \\
$T_3-$(T) haVTZ(f,p) 3-body VMFC& -0.007 & & & & & +0.008 \\
$T_3-$(T) VTZ(f,p) 3-body noCP & 0.022 & 0.020 & 0.016 & 0.016 & 0.016 & 0.011 &0.011 &0.012 & 0.041\\
$T_3-$(T) VTZ(f,p) 3-body VMFC &&&&&&&&&0.000\\
$T_3-$(T) V\{D,T\}Z 3-body& 0.026& 0.022& 0.019& 0.019& 0.019& 0.015& 0.015& 0.016\\
$T_3-$(T) VDZ 4-body noCP & 0.005 & 0.003 & 0.003 & 0.004 & 0.003 & 0.002 & 0.002 & 0.002 \\
$T_3-$(T) VDZ 4-body noCP & 0.006 \\
$T_3-$(T) VDZ 4-body VMFC & 0.002 \\
$T_3-$(T) VDZ 5+6-body noCP & 0.000 & 0.000 & 0.000 & 0.000 & 0.001 & 0.001 & 0.001 & 0.001\\
\hline
(Q) VTZ(f,p) 2-body noCP &0.173 & 0.167 & 0.139 & 0.141 & 0.143 & 0.107 & 0.108 & 0.107\\
(Q) VTZ(f,p) 2-body VMFC &0.101 & 0.100 & 0.086 & 0.088 & 0.087 & 0.067 & 0.068 & 0.067\\
(Q) VQZ(g,d) 2-body noCP & +0.200 & & & & & +0.126 \\
(Q) VQZ(g,d) 2-body VMFC & +0.133 & & & & & +0.086 \\
(Q) haVTZ(f,p) 2-body noCP& +0.160 & +0.155 & +0.134 & +0.135 & +0.135& +0.106 & +0.106& +0.105 &+0.243 \\
(Q) haVTZ(f,p) 2-body VMFC& +0.141 & +0.137 & +0.117 & +0.119 & +0.118& +0.092 & +0.092 & +0.091&+0.214 \\
(Q) haVQZ(g,d) 2-body noCP& +0.175 &+0.170 &+0.145 &+0.147 &+0.147 & +0.115 &+0.115&+0.114& +0.265 \\
(Q) haVQZ(g,d) 2-body VMFC& +0.163 &+0.158 &+0.135 & +0.137 &+0.137 & +0.105 &+0.106&+0.105&+0.247 \\
(Q) haV\{T,Q\}Z 2-body noCP&0.18&0.18&0.15&0.15&0.15&0.12&0.12&0.12&0.28\\
(Q) haV\{T,Q\}Z 2-body VMFC&0.18&0.17&0.15&0.15&0.15&0.11&0.11&0.11&0.27\\
(Q) best estimate & 0.18$\pm$0.004&0.17$\pm$0.004&0.15$\pm$0.003&0.15$\pm$0.003&0.15$\pm$0.003&0.12$\pm$0.004&0.12$\pm$0.003&0.12$\pm$0.003&0.27$\pm$0.01\\
(Q) haVTZ(f,p) 3-body noCP& +0.011 & & & & & +0.004 \\
(Q) haVTZ(f,p) 3-body VMFC& +0.006 & & & & & +0.004 \\
(Q) VTZ(d,p) 3-body noCP & +0.017 & & & & & +0.008 & & &   +0.037\\
(Q) VTZ(d,p) 3-body VMFC & +0.008 & & & & & +0.005 & & &   +0.020\\
(Q) VTZ(f,p) 3-body noCP &  & & & & &  & & & 0.035\\
(Q) VTZ(f,p) 3-body VMFC &  & & & & &  & & & 0.020\\
(Q) 4-body (a) noCP & 0.015 &0.012&0.012&0.012&0.013&0.011&0.010&0.010\\
(Q) 4-body VDZ noCP & 0.012 \\
(Q) 4-body VDZ VMFC & 0.003\\ 
(Q) 5+6body noCP (a) & -0.001 & 0.000 & 0.001 & 0.001 & 0.001 & 0.003 & 0.003 & 0.003\\
\hline
$T_4-(Q)$ 2-body VDZ(d,s)&-0.034&-0.034&-0.028&-0.028&-0.028&-0.021&-0.021&-0.021\\
$T_4-(Q)$ 2-body VTZ(d,p)&-0.029&-0.028&-0.020&-0.021&-0.022&-0.014&-0.014&-0.014\\
$T_4-(Q)$ 2-body
haVDZ noCP&-0.010&&&&&&&&-0.016\\
$T_4-(Q)$ 2-body
haVDZ VMFC&-0.007&&&&&&&&-0.011\\
$T_4-(Q)$ 2-body VTZ(f,p)&-0.024&&&&&\\
$(5)$ 2-body VDZ(d,s) noCP& 0.014 & 0.014 & 0.013 & 0.013 & 0.013 & 0.011 &0.011 &0.011&0.021\\
$(5)$ 2-body VDZ(d,s) VMFC&0.006
&&&&&&&&0.009\\\hline
CV 
2-body CCSD(T)/awCV\{T,Q\}Z& 0.306 & 0.310 & 0.326 & 0.323 & 0.319 & 0.344 & 0.334 & 0.336\\
ditto 3-body & 0.009 & 0.008 & 0.013 & 0.010 & 0.011 & 
0.021 & 0.018 & 0.020\\
CV CCSD(T)/hawCVTZ 2-body noCP& 0.308 &0.311 &0.322 &0.320 & 0.317 & 0.332 & 0.324  & 0.325  & 0.466 \\
CV CCSD(T)/hawCVTZ 2-body VMFC& 0.276 &0.278 &0.290&0.288&0.284& 0.302 & 0.294  & 0.296  & 0.414 \\
CV CCSD(T)/hawCVQZ 2-body noCP& 0.309 &0.312 &0.325&0.323&0.320& 0.340 & 0.331  &0.332   & 0.464 \\
CV CCSD(T)/hawCVQZ 2-body VMFC& 0.304 &0.307&0.321&0.318&0.315& 0.335 & 0.327  & 0.328  & 0.457 \\
CV CCSD(T)/hawCV5Z 2-body noCP& 0.314 \\
CV CCSD(T)/hawCV5Z 2-body VMFC& 0.312 \\
CV CCSD(T)/hawCVTZ 3-body noCP& 0.009\\
CV CCSD(T)/hawCVTZ 3-body VMFC& 0.014\\
CV CCSD(T)/hawCVQZ 3-body noCP& 0.013\\
CV CCSD(T)/hawCVQZ 3-body VMFC& 0.015\\
hawCV\{T,Q\}Z 2-body noCP &0.309&0.312&0.328&0.325&0.322&0.345&0.336&0.337&0.463\\
hawCV\{T,Q\}Z 2-body VMFC &0.324&0.328&0.343&0.341&0.337&0.360&0.350&0.352&0.488\\
CV best estimate & 0.317$\pm$0.008& 0.320$\pm$0.008&0.336$\pm$0.008&0.333$\pm$0.008&0.329$\pm$0.008&0.352$\pm$0.007&0.343$\pm$0.007&0.345$\pm$0.007&0.476$\pm$0.013\\
hawCV\{Q,5\}Z 2-body noCP & 0.319 \\
hawCV\{Q,5\}Z 2-body VMFC & 0.320 \\
REL MP2/AVTZuc & -0.097 &
-0.093 &
-0.098 & 
-0.096 & 
-0.094 &
-0.106 &
-0.102 &
-0.103\\
ditto 3-body & -0.007 & -0.005& -0.008 & -0.006 &-0.008 & -0.014 & -0.012 & -0.014\\
ditto 4-body & 0.002 &0.002& 0.002 & 0.002&0.002&0.002&0.002&0.002\\
REL CCSD(T)/AVTZuc 2-body & -0.102&-0.098&-0.103&-0.101&-0.099&-0.111&-0.107&-0.109&-0.140\\
DBOC (b) & 0.130 & 0.130 & 0.129 & 0.129 & 0.126 & 0.126 & 0.122 & 0.123 & 0.195 \\
\hline\hline
\end{tabular}
}

\begin{footnotesize}
(a) CCSDT(Q)/cc-pVDZ scaled\cite{jmlm200} by 1.25 as in W4lite;
 (b) CCSD/AVTZ 2-body; 3-body and 4-body terms vanish to all decimal places.
\end{footnotesize}
\end{table*}

Results for the eight \ce{(H2O)6} isomers are gathered in Table~\ref{tab:water6}, while those for the four Wales-Hodges \ce{(H2O)20} structures can be found in Table~\ref{tab:water20}.

In line with what is seen for general thermochemistry,\cite{jmlm200,jmlm273}
the two-body higher-order triples corrections, $T_3$--(T), are universally antibonding, and this trend is seen both with and without VMFC, as well as for both cc-pVTZ(f,p) and haVTZ(f,p) basis sets. 

The 3-body terms, in contrast, appear to be bonding at first sight, but when VMFC is applied, they seem to largely vanish. 

Expanding the basis set from VTZ(f,p) to VQZ(g,d) causes a small increase (in absolute value) in 
$T_3$--(T).
From haVTZ(f,p) to haVQZ(g,d), there is less of a basis set effect even without CP correction, and almost none with VMFC. haV\{T,Q\}Z extrapolation according to Ref.\citenum{Karton2020} still leaves a small gap between noCP and VMFC estimated basis set limits. It was argued by Sherrill and coworkers\cite{Burns2014a} (see also Brauer et al.\cite{jmlm259}) that for orbital-based CCSD(T) calculations with medium-sized basis sets, the noCP and CP-corrected noncovalent dissociation energies bracket the basis set limit. We hence derive the `best estimate' values as the average of the noCP and VMFC extrapolated values, and the uncertainty interval as one-half the difference (i.e., stretching to both noCP and VMFC values).

The differential counterpoise corrections on the higher-order triples correction behave anomalously. That is, they actually increase as the basis set is expanded from haVTZ to haVQZ, even though the \emph{overall} CP correction is reduced by more than half. Note also that we see no such anomaly for the differential CP on the connected quadruples contribution.

In a recent paper\cite{jmlm347},
 we explored the convergence behavior of post-CCSD(T) contributions to molecular atomization energies with basis sets through septuple-zeta. We found that the slow and somewhat erratic basis set convergence of CCSDT $-$ CCSD(T) is actually a superposition of countervailing trends: a relatively rapid convergence of CCSDT-3\cite{Noga1987} $-$ CCSD(T) and a very slow convergence of CCSDT $-$ CCSDT-3, which primarily stems from the direct impact of triple excitations on the triples amplitudes, starting at $E^{(5)}_{TT}$ in perturbation theory.\cite{Cremer2001} For quintuple-zeta and larger basis sets, this behavior dominates the entire higher-order triples contribution and is smooth.

 Clearly, using (nonhydrogen-)augmented quintuple zeta basis sets is not a computationally viable option at present for \ce{(H2O)6}, let alone for \ce{(H2O)20}. But as a probe, we singled out the strongest pair interaction in the \ce{(H2O)6} prism structure, which happens to be the 1--5 interaction in the geometry as given. The results are given in Table~\ref{tab:pair15}.

 \begin{table}[h!]
\caption{Post-CCSD(T) contributions to $D_e$ (cal/mol; positive values increase binding) for the strongest hydrogen-bonded pair (waters 1--5) of the (H$_2$O)$_6$ prism. 
$\Delta$BSSE = noCP $-$ VMFC. 
For scale, the CCSD(T) 2-body $D_e$ (noCP/VMFC) is 3.988/3.425, 4.115/3.873, and 4.096/4.000~kcal/mol in the haVTZ(f,p), haVQZ(g,d), haV5Z(h,f) basis sets.\label{tab:pair15}}
\centering
\footnotesize
\begin{tabular}{lrrr}\hline
 & noCP & VMFC & $\Delta$CP \\
\hline
\multicolumn{4}{l}{$[\mathrm{CCSDT\text{-}3}]-[\mathrm{CCSD(T)}]$}\\
haVTZ & $-$24.0 & $-$28.3 & +4.2 \\
haVQZ & $-$24.8 & $-$27.1 & +2.3 \\
haV5Z & $-$25.7 & $-$26.4 & +0.7 \\
\multicolumn{4}{l}{$[\mathrm{CCSDT}]-[\mathrm{CCSDT\text{-}3}]$}\\
haVTZ & +7.2 & +13.0 & $-$5.8 \\
haVQZ & +4.8 & +10.4 & $-$5.6 \\
haV5Z & +4.7 & +7.9 & $-$3.1 \\
\multicolumn{4}{l}{$T_3-$(T)}\\
haVTZ & $-$16.9 & $-$15.3 & $-$1.6 \\
haVQZ & $-$20.0 & $-$16.7 & $-$3.3 \\
haV5Z & $-$20.9 & $-$18.5 & $-$2.4 \\
haV\{T,Q\}Z$^a$ & $-$23.0 & $-$18.1 & $-$4.9 \\
haV\{T,Q\}Z$^b$ & $-$24.0 & $-$18.6 & $-$5.4 \\
haV\{Q,5\}Z$^c$ & $-$22.2 & $-$21.1 & $-$1.1 \\
\multicolumn{4}{l}{(Q)}\\
haVTZ & +23.5 & +21.2 & +2.3 \\
haVQZ & +25.4 & +23.8 & +1.6 \\
haV\{T,Q\}Z$^d$ & +26.6 & +25.5 & +1.1 \\
\hline\hline
\end{tabular}

\raggedright\footnotesize
$^a$$\alpha\approx2.5$, as in Table~3 of Karton\cite{Karton2020}. $^b$$\alpha=2.012$; $^c$$\alpha=2.35$; $^d$$\alpha=3.24$; all 3 taken from Ref.\citenum{jmlm347}.
\end{table}

 Note that $\Delta\mathrm{CP}$ for CCSDT-3 $-$ CCSD(T) behaves as expected across all three basis sets, while $\Delta\mathrm{CP}$ for CCSDT $-$ CCSDT-3 stagnates across haVTZ(f,p) and haVQZ(g,d) and only decays when one proceeds further to haV5Z(h,f). Adding the two up (note the opposite signs of the components) reproduces the counterintuitive behavior for haVTZ(f,p) and haVQZ(g,d), and reveals that haV5Z(h,f) again behaves intuitively.

  We are therefore confident that our basis set limit extrapolations actually do reflect reality, and in fact suspect that our uncertainty bars, which are based on the difference between raw and counterpoise, may actually be somewhat conservative.

Let us turn now to (Q), which --- as expected from thermochemical experience --- is bonding across the board. Here we see a sizable VMFC correction for the VTZ(f,p) basis set, across all isomers, and this even persists for the VQZ(g,d) results we were able to obtain for the prism (Iso1 PR) and chair (Iso6 CC) isomers. In contrast, haVTZ(f,p) counterpoise corrections are a factor of 3--4 smaller (more like a factor of 6 at the CCSD(T) level), yielding values midway between `raw' and CP-corrected VTZ(f,p). Expanding the basis set to haVQZ(g,d) causes a modest change in the `raw' (Q) correction, but the counterpoise correction is reduced from 0.019 to 0.012 kcal/mol for the prism. 
Unlike for the higher-order triples, for (Q)/haV\{T,Q\}Z there is only a fairly small gap between noCP and VMFC estimated basis set limits, and hence the `best estimates' only carry small uncertainties. It has been known for some time (e.g., Ref.\citenum{jmlm205}) that higher-order triples converge more slowly with the basis set than connected quadruples.

Again, without CP correction we find some small (Q) 3-body contributions, 
and even these are halved
when VMFC is applied. This appears to justify treating post-CCSD(T) contributions in larger clusters like \ce{(H2O)20} in the 2-body approximation, albeit with 2-body VMFC.

For \ce{(H2O)6}, we find the four-body contribution to be altogether insignificant, especially with VMFC correction: $T_3-$(T) 0.002 and (Q) 0.003 kcal/mol. We hence shall make no attempt to evaluate it for the water icosamers.

Let us now compare two-body post-CCSD(T) contributions between isomers. While the higher-order triples contribution is relatively constant across the series, (Q) clearly favors more compact structures like the prism (Iso1) over more open structures like the chair (Iso6). This is probably most easily understood in terms of the former having more close contacts than the latter: for example, one can easily see (Figure~\ref{fig:waterclusters}) that Iso1 has nine hydrogen `bridges', compared to just six for Iso6.

As an additional control, we repeated the calculations for the cubic conformer of \ce{(H2O)8}. The trends seen for the \ce{(H2O)6} prism are observed here on a somewhat greater scale, namely: mutual cancellation of post-CCSDT(Q) correlation corrections, a binding (Q) correction that now reaches 1 kJ/mol (0.24 kcal/mol) for the haVTZ basis set and exceeds that figure upon extrapolation, and an antibonding higher-order triples contribution about half the size of the (Q) contribution.

A noteworthy feature across haVTZ, haVQZ, and extrapolated haV\{T,Q\}Z is that the total post-CCSD(T) correction, i.e., CCSDT(Q) $-$ CCSD(T) or the sum of $T_3-(T)$ and (Q), is remarkably stable across basis sets. 
One becomes tempted to just use haVTZ for the larger clusters.

As seen in general thermochemistry\cite{jmlm200,jmlm205}, the higher-order connected quadruples term $T_4 - (Q)$, i.e. the difference between fully iterative CCSDTQ\cite{Kucharski1992} and CCSDT(Q), is slightly antibonding in the absence of counterpoise. However, with the small cc-pVDZ basis set that lacks diffuse functions, this correction is somewhat exaggerated, and upon VMFC correction it drops down to almost nil.
Inclusion of connected quintuples,\cite{mrcc63,mrcc65} as shown previously\cite{jmlm200,jmlm205,jmlm340} for general thermochemistry, partly cancels this contribution, being binding by 0.014 and 0.011 kcal/mol, respectively, for prism and ring. (For this step, we used the cc-pVDZ(d,s) basis set, which was previously found\cite{jmlm200,jmlm205,jmlm340} to be more than adequate for this high excitation contribution, at least in a thermochemical context.) Both $T_4 - (Q)$ and $(5)$ corrections are significantly (but not fully) reduced when VMFC is applied, leaving 
their sum to be almost nil. Hence,
we deem it both expedient and acceptable for the \ce{(H2O)20} isomers to ignore post-CCSDT(Q) correlation corrections.

\subsection{Post-CCSD(T) effects in \ce{(H2O)20} isomers}

\begin{table*}[h!]

\caption{Dissociation energy contributions for \ce{(H2O)20} Wales-Hodges isomers (kcal/mol)\label{tab:water20}}
\centering
\resizebox{0.9\textwidth}{!}{
\begin{tabular}{lrrrr}\hline
& ES & FC & FS & \\
& edge-sharing & fused cubes & face-sharing & dodecahedron \\
& pentagons & (box-kite) & stacked pentagons &\\
\hline
&\multicolumn{4}{c}{valence CCSD(F12*)/CBS +(T) limit for perspective only}\\
Ref.\citenum{jmlm272}& 219.30 & 216.31 & 217.58 & 211.22\\
\hline
$T_3-$(T) VTZ(f,p) 2-body noCP&-0.431\\
$T_3-$(T) VTZ(f,p) 2-body VMFC&-0.367\\
$T_3-$(T) VQZ(g,d) 2-body noCP&-0.539\\
$T_3-$(T) VQZ(g,d) 2-body VMFC&-0.391\\
$T_3-$(T) haVTZ(f,p) 2-body noCP&-0.299&-0.284&-0.294&-0.315 \\
$T_3-$(T) haVTZ(f,p) 2-body VMFC&-0.254&-0.235&-0.248&-0.277 \\
$T_3-$(T) haVQZ(g,d) 2-body noCP&-0.379&-0.359&-0.373&-0.398 \\
$T_3-$(T) haVQZ(g,d) 2-body VMFC&-0.284&-0.263&-0.278&-0.311 \\
$T_3-$(T) haV\{T,Q\}Z 2-body noCP&-0.46&-0.43&-0.45&-0.48 \\
$T_3-$(T) haV\{T,Q\}Z 2-body VMFC&-0.31&-0.29&-0.31&-0.34 \\
$T_3-$(T) best estimate & -0.38$\pm$0.07 &-0.36$\pm$0.07&-0.38$\pm$0.07& -0.41$\pm$0.07 \\
(Q) VTZ(f,p) 2-body noCP&+0.714\\
(Q) VTZ(f,p) 2-body VMFC&+0.428\\
(Q) VQZ(g,d) 2-body noCP&+0.839\\
(Q) VQZ(g,d) 2-body VMFC&+0.560\\
(Q) haVTZ(f,p) 2-body noCP&+0.680&0.714&0.684&0.599 \\
(Q) haVTZ(f,p) 2-body VMFC&+0.594&0.626&0.596&0.521 \\
(Q) haVQZ(g,d) 2-body noCP&+0.748&+0.786&+0.753&+0.652 \\
(Q) haVQZ(g,d) 2-body VMFC&+0.692&+0.731&+0.697&+0.601 \\
(Q) haV\{T,Q\}Z 2-body noCP&+0.79&+0.83&+0.79&+0.68\\
(Q) haV\{T,Q\}Z 2-body VMFC&+0.75&+0.79&+0.76&+0.65\\
(Q) best estimate & +0.77$\pm$0.02 & +0.81$\pm$0.02   & +0.77$\pm$0.02   & +0.67$\pm$0.02\\
CV CCSD(T)/hawCVTZ 2-body noCP& 1.367 & 1.291 & 1.339 & 1.362 \\
CV CCSD(T)/hawCVTZ 2-body VMFC& 1.214 & 1.140 & 1.187 & 1.213 \\
CV CCSD(T)/hawCVQZ 2-body noCP& 1.361 &1.276&1.330& 1.363 \\
CV CCSD(T)/hawCVQZ 2-body VMFC& 1.339 &1.255&1.308& 1.343 \\
CV CCSD(T)/hawCV\{T,Q\}Z 2-body noCP& 1.357 &1.266&1.322& 1.364 \\
CV CCSD(T)/hawCV\{T,Q\}Z 2-body VMFC& 1.430 &1.339&1.397& 1.438 \\
CV best estimate 2-body & 1.39$\pm$0.04 & 1.30$\pm$0.04 & 1.36$\pm$0.04 & 1.40$\pm$0.04 \\
scalar relativistics 2-body CCSD(T)/haVTZ(uncon.)& -0.43 & -0.41 & -0.42 & -0.41\\
DBOC 2-body CCSD/AVDZ& 0.57 & 0.54 & 0.56 & 0.55\\
\hline\hline
\end{tabular}
}

\begin{footnotesize}
Reference geometries taken unaltered from WATER27 dataset, via GMTKN55
\end{footnotesize}

\end{table*}

Table \ref{tab:water20} shows the various contributions. Turning first to the haVTZ(f,p) results, we again see that higher-order triples have very similar contributions across all four Wales-Hodges isomers. The connected quadruples, however, seem to be less favorable for the most open, dodecahedral, isomer than for the other three, and among the latter, the box-kite (four stacked cubes) appears to have a slight edge over the two other structures --- the edge-sharing pentagons and the stacked pentagons.

While the (Q) do not change the ordering of the isomers, a contribution on the order of 0.7--0.8 kcal/mol is quite nontrivial in accurate thermochemical work. The boost over the dodecahedron that the more compact structures receive from quadruples may be small in the greater scheme of things, but does reveal an interesting detail about isomerism in water clusters --- one that is present both in smaller clusters like \ce{(H2O)6} and in larger ones like \ce{(H2O)20}.

With the VTZ(f,p) basis set, the VMFC counterpoise correction on (Q)  is almost 0.3 kcal/mol for the edge-shared isomer. However, for haVTZ this gap shrinks to 0.09 kcal/mol, and for haVQZ(g,d) further to 0.06 kcal/mol. 
In fact, like for \ce{(H2O)6}, the total post-CCSD(T) contribution is relatively stable with the basis set, settling at 0.33 (noCP) and 0.44 (VMFC) kcal/mol upon basis set extrapolation.
In other words: the post-CCSD(T) contribution \emph{is quite genuine and cannot be reasonably dismissed as an artifact of BSSE}.



This picture would be somewhat more ambiguous if un-augmented cc-pVTZ and cc-pVQZ basis sets were considered.



(Q) is almost perfectly proportional to the number of H-bonds (HBs) in the cluster: 0.022-0.023 kcal/mol per H-bond for the icosamers (which have 30 HBs for the dodecahedron, 36 for fused cubes, 35 for face-sharing and 34 for edge-sharing pentagonal prisms\cite{Xantheas2004}) and 0.020--0.021 kcal/mol per H-bond for the hexamers (6 HBs for the ring, 9 HBs for the prism).

\subsection{Subvalence correlation and further corrections}
 
For molecules without significant static correlation, the fairly systematic mutual cancellation between bonding quadruples and largely antibonding higher-order triples is what causes CCSD(T) to be a `Pauling point' (see also Stanton\cite{Sta97} for a perspective based on perturbation theory). While this cancellation is broadly conserved here as well,
we have seen that higher-order correlation corrections subtly but unambiguously favor the more compact structures.

It then, however, behooves us to investigate other potential sources of differences.

Scalar relativistic corrections reduce the dissociation energies of the \ce{(H2O)6} isomers across the board by about 0.1 kcal/mol, and those of the four \ce{(H2O)20} isomers by about 0.4 kcal/mol. This is fortuitously somewhat over-compensated by the CCSD/AVDZ diagonal Born-Oppenheimer corrections, which {\em increase} the dissociation energies across the board by about 0.12--0.13 kcal/mol for \ce{(H2O)6} and by 0.56$\pm$0.01 kcal/mol for \ce{(H2O)20}. Both terms are fairly consistent between isomers.

The 3-body contribution to the scalar relativistic correction is an order of magnitude less than the two-body term, while the four-body term is altogether insignificant. The 3-body and 4-body DBOC terms vanish entirely.

The core-valence correlation is another matter. Cognizant of the impact of BSSE on core-valence contributions\cite{LeeRendellTaylor1990,BauschlicherRicca1998}, we evaluated these contributions with and without VMFC, at the CCSD(T) level with hawCVTZ and hawCVQZ basis sets. As can be seen in Tables~\ref{tab:water6} and \ref{tab:water20},  differences between noCP and VMFC with the hawCVTZ basis set reach about 0.15 kcal/mol for \ce{(H2O)20}, but this shrinks about sevenfold upon proceeding to the hawCVQZ basis set. The \{T,Q\} extrapolated values differ slightly between noCP and VMFC; as above for $T_3-(T)$ and $(Q)$, we are taking the average as the best estimate and one-half the difference as the uncertainty. For the prism structure of \ce{(H2O)6}, we additionally considered the hawCV5Z basis set and the \{Q,5\} limit: here the noCP and VMFC results are essentially the same, and quite close to the abovementioned average.

For the \ce{(H2O)6} isomers, the best estimates settle around 0.32--0.35 kcal/mol, with uncertainties of about 0.01 kcal/mol. For \ce{(H2O)8}, this increases to 0.48$\pm$0.01 kcal/mol, while for the \ce{(H2O)20} isomers, we reach 1.30--1.40 kcal/mol, with uncertainties of 0.04 kcal/mol on the respective values. In fact, this makes 2-body core-valence corrections {\em the} most important term beyond the valence CCSD(T) limit.


This flies in the face of the  received wisdom that core-valence (CV) correlation does not matter for noncovalent interactions. While we have evaluated it in high-accuracy work on benzene dimer and benzene...ethylene,\cite{jmlm298} it was found to be insignificant there. It would stand to reason that this would be the case for distant contacts; however, for closer contacts like hydrogen bonds, CV correlation can apparently not be ignored. Boese\cite{Boese2013} and \v{R}ez\'a\v{c}, Hobza, and coworkers\cite{A24bis} previously found 0.03 kcal/mol for water dimer: the values found presently are comparable to, if somewhat larger than, what one might expect from scaling this contribution by the number of hydrogen-bond contacts. 

Intriguingly, CV is larger for the more open structures than for the more compact ones: chair 0.352 vs.\ prism 0.317 kcal/mol for \ce{(H2O)6}, and dodecahedron 1.40 vs. fused cubes 1.30 kcal/mol. (The slight 3-body CV is also somewhat larger for the cyclic hexamers.) This actually runs counter to the trend of the valence post-CCSD(T) corrections. This can be rationalized by analyzing the O--H\ldots O contact distances: in the hexamer ring, for example, the six such contacts have similar distances of about 1.73 \AA, while for the prism, only the three inter-triangle distances are short (ca. 1.67 \AA) while the six intra-triangle distances are as long as 2.17 \AA. 

Three-body core-valence correction effects are much less significant: for the prism hexamer, we obtain 0.013 (noCP) or 0.015 (VMFC) kcal/mol using the hawCVQZ basis set. Reducing the basis set to hawCVTZ, the VMFC value is fairly stable (0.014 kcal/mol), but the noCP correction drops to 0.009 kcal/mol. Depending on the structure, we have 3-body CV corrections that are 1--1.5 orders of magnitude smaller than the two-body terms.

\section{Conclusions}

While for water clusters, post-CCSD(T) valence correlation contributions cancel to some degree between antibonding higher-order triples and connected quadruples, the cancellation is neither perfect nor homogeneous: connected quadruples favor more compact structures (such as the water hexamer prism) over more spread-out ones (particularly rings). Overall, post-CCSD(T) valence correlation enhances the binding energy. In the absence of counterpoise corrections, artifactual 3-body and 4-body corrections may occur, especially with basis sets lacking in diffuse functions. Valence post-CCSD(T) three-body contributions largely vanish when VMFC (Valiron-Mayer function counterpoise) corrections are applied, and four-body post-CCSD(T) entirely so.

Relativistic corrections and DBOC fortuitously largely cancel each other. The most significant contribution for water clusters beyond  valence CCSD(T) is actually core-valence correlation, followed by connected quadruples. While core-valence correlation tends to favor more open structures, (Q) is roughly proportional to the number of hydrogen bonds and thus favors more compact structures. 


Finally, we can state that the bulk  ($\geq$94\%) of contributions beyond valence CCSD(T) is recovered by only 2-body terms. This is what makes their consideration for larger water clusters a viable proposition at all.

\begin{acknowledgement}
Research at Weizmann was supported by the Israel Science Foundation (grant 1969/20), by the Minerva Foundation (grant 2020/05), and 
by an internal grant from the Tom and Mary Beck Center for Advanced and Intelligent Materials (C-AIM) fund. 
ES thanks the Weizmann School of Science (WISsE) for a doctoral fellowship and the Onassis Foundation (Scholarship ID: FZP 052-2/2021-2022), while MS acknowledges an excellence postdoctoral fellowship from WISsE.
Computing facilities were provided by the Faculty of Chemistry HPC facility CHEMFARM, the maintenance of which is funded by the Ben May Center for Chemical Theory and Computation (Weizmann Institute of Science).

This research was originally inspired by discussions with Prof. Kieron Burke (UC Irvine) about the \ce{(H2O)6} isomers. We additionally thank Profs. A. Daniel Boese (University of Graz, Austria) and Leslie Leiserowitz (Weizmann Institute) for helpful discussions, and Prof. Devin A. Matthews (Southern Methodist University, Dallas, Texas), Dr. Gregory Jones (Quantum Theory Project, University of Florida) and QTP's late lamented chair, Prof. John F. Stanton (1961--2025) 
for assistance with CFOUR and NCC.
\end{acknowledgement}

\begin{suppinfo}

\begin{itemize}
\item Microsoft Excel workbook containing the full raw data.
\item Python scripts for setting up and processing VMFC calculations
\end{itemize}

\end{suppinfo}

\footnotesize
\bibliography{waterNccsdtq}%
\end{document}